\documentclass[conference]{IEEEtran}
\IEEEoverridecommandlockouts

\usepackage{dblfloatfix}
\usepackage{comment}
\usepackage{cite}
\usepackage{amsmath,amssymb,amsfonts}
\usepackage{algorithmic}
\usepackage{graphicx}
\usepackage{textcomp}
\usepackage{xcolor}
\def\BibTeX{{\rm B\kern-.05em{\sc i\kern-.025em b}\kern-.08em
    T\kern-.1667em\lower.7ex\hbox{E}\kern-.125emX}}
\begin{document}


\title{Self-Adaptive Microservice-based Systems - Landscape and Research Opportunities
\\
}


\author{\IEEEauthorblockN{Messias Filho, Eliaquim Pimentel, Wellington Pereira, Paulo Henrique M. Maia, Mariela I. Cortés}
\IEEEauthorblockA{\textit{State University of Ceará (UECE)} \\
Fortaleza, Ceará, Brazil \\
\{messias.filho, eliaquim.pimentel\}@aluno.uece.br, \{wellington.costa128\}@gmail.com, \\ \{pauloh.maia, mariela.cortes\}@uece.br}
}


\maketitle


\begin{abstract}

Microservices have become popular in the past few years, attracting the interest of both academia and industry. Despite of its benefits, this new architectural style still poses important challenges, such as resilience, performance and evolution. Self-adaptation techniques have been applied recently as an alternative to solve or mitigate those problems. However, due to the range of quality attributes that affect microservice architectures, many different self-adaptation strategies can be used. Thus, to understand the state-of-the-art of the use of self-adaptation techniques and mechanisms in microservice-based systems, this work conducted a systematic mapping, in which 21 primary studies were analyzed considering qualitative and quantitative research questions. The results show that most studies focus on the Monitor phase (28.57\%) of the adaptation control loop, address the self-healing property (23.81\%), apply a reactive adaptation strategy (80.95\%) in the system infrastructure level (47.62\%) and use a centralized approach (38.10\%). From those, it was possible to propose some research directions to fill existing gaps.  

\end{abstract}

\begin{IEEEkeywords}

Microservice architecture, Self-adaptation, Systematic study mapping, Software engineering

\end{IEEEkeywords}

\section{Introduction}
\label{sec:introduction}

In software engineering, architectural design aims to provide a bridge between system functionality and the quality requirements that the system must meet. Over the past decades, software architecture has been studied thoroughly and, as a result, software engineers have created different ways of composing systems that provide broad functionality and satisfy a variety of requirements~\cite{dragoni2016}.

In this realm, the microservice architectural style arises as a strategy to develop a single application as a set of small services, each running its own process, and communicating through lightweight mechanisms, usually a HTTP resource API~\cite{lewis2014}. It is a technology-agnostic approach, which means that a microservice can be implemented regardless of the programming language used, as long as it provides its functionality via message exchange~\cite{dragoni2016}.

The popularity of this architectural style has grown boosted by its adoption by several large companies such as Netflix, Amazon, Linkedin among others~\cite{lewis2014}~\cite{villamizar2015}. It has also attracted the interest of the academic community and other areas of research, standing out for its use in areas such as blockchain~\cite{Sousa2020}, Internet of Things (IoT)~\cite{sun2017} and mobile systems~\cite{fan2017}. Microservices are also appropriate for systems deployed in cloud infrastructures, promoting greater utilization of the elasticity resources and on-demand provisioning of the cloud model~\cite{alshuqayran2016}.

Despite its benefits, as the number of deployed microservices increases so does its complexity, bringing some challenges to this architectural style~\cite{shankararaman2020}. Therefore, some approaches are necessary to monitor the microservices' status (and the containers where they execute) at runtime in order to provide transparency, elasticity, resilience and deployment management.

One approach that has been used to solve or mitigate those problems is the use of self-adaptation, which allows the system to adjust its behavior or structure at runtime in response to environmental changes~\cite{cheng2009} \cite{lemos2013}. Thus, a self-adaptive system must continuously monitor changes in its context and react accordingly~\cite{cheng2009}.

This paper aims at investigating and exploring the use of self-adaptation techniques and mechanisms in the microservice architectural style in order to understand how those concepts have been used to improve microservice-based systems. To do that, we conducted a systematic literature mapping~\cite{kitchenham2011} that contemplates 10 primary research questions (one qualitative and the other quantitative questions) answered by 21 studies taken from five digital libraries. Our main contribution consists of the analysis and discussion of the state-of-the-art of the studied subject, thus helping researchers and practitioners (i) to understand how  microservices are currently benefited by the use of self-adaptive techniques such that they can possibly reuse and extend such approaches in their own work, and (ii) to address current limitations of those studies by new research directions.

The rest of this paper is organized as follows: Section \ref{sec:background} presents a brief overview of microservice-based architecture. The systematic mapping is detailed in Section \ref{sec:systematic-mapping-study}, while Section \ref{sec:discussion} discusses the obtained results and provides research directions. Threats to the validity of this work are addressed in Section \ref{sec:threats-to-validaty}. Finally, we draw our conclusions and future work in  Section \ref{sec:conclusion-and-future-works}.

\section{Microservice-based architecture}
\label{sec:background}

Microservices are an architecture style in which small and autonomous services \cite{newman2015} run in its own process and communicate to each other with lightweight mechanisms, often an HTTP resource API \cite{lewis2014}. In this way, microservices tackles the increase in complexity by decomposing the functionality of a large system into a set of independent services, thus emphasizing modularity through loose coupling and high cohesion \cite{dragoni2016}.

Those services need to be able to evolve independently of each other and be deployed without system reboot. Among the main benefits of microservices we can cite the technology heterogeneity, resilience, ease of deployment and maintainability \cite{newman2015}. Furthermore, as each microservice can be managed independently, this architectural style  proposes a solution to scale computational resources efficiently in cloud environments. For this reason, the microservice-based architecture  has been used as an alternative to traditional monolithic architecture \cite{villamizar2015} and many studied have proposed evolution approaches to reach this new paradigm~\cite{Barbosa2020}. The main layers involved in this architectural style are the application layer, which involves aspects of business and system functionality, and the infrastructure layer, which deals with resource management and structural functioning of the system.
\section{Systematic Mapping Study}
\label{sec:systematic-mapping-study}

\begin{table}[]
\caption{Research questions}
\label{tab:tableresearch}
\begin{tabular}{p{0.9cm} p{6.5cm}}
\hline
\textbf{ID} & \textbf{Research Question} \\ \hline
RQ1 & How has self-adaptation been applied in the context of microservices? \\
RQ2 & What types of research and contribution have been presented? \\
RQ3 & Which phase of a self-adaptation control loop is the focus of the studies?  \\
RQ4 & What self-* properties have been addressed? \\
RQ5 & What self-adaptation strategies have been used? \\
RQ6 & What quality requirements have been addressed by the approaches? \\
RQ7 & In which microservice architecture layer were the adaptations applied? \\
RQ8 & What self-adaptation control logic has been addressed? \\
RQ9 & What technologies have been addressed? \\
RQ10 & Has an empirical evaluation been conducted? \\
RQ10.1 & What strategy was used to validate the research? \\
\hline
\end{tabular}
\end{table}

\begin{table*}[]
\centering
\caption{Search String}
\label{tab:tablesearchstring}
\begin{tabular}{p{2.0cm} p{15cm}}
\hline
\textbf{Group} & \textbf{String} \\ \hline
Microservice  & (``microservice'' OR ``micro service'' OR ``micro-service'') \\
Self-adaptation & (``self*'' OR ``runtime adaptation'' OR ``runtime adaptive'' OR ``adaptive software'' OR ``adaptive systems'' OR ``adaptive system'' OR ``adaptive services'' OR ``adaptive service'' OR ``adaptable software'' OR ``adaptability'' OR ``adaptivity'' OR ``context-aware'' OR ``context-awareness'' OR ``dynamic updates'' OR ``dynamic adaptation'' OR ``dynamic system'' OR ``dynamic systems'' OR ``autonomous systems'' OR ``autonomous software'' OR ``autonomic computing'' OR ``runtime reconfiguration'') \\ \hline
\end{tabular}
\end{table*}

This study follows the guidelines established by Kitchenham and Charters~\cite{kitchenham2007, kitchenhambook2015} and applies the mapping process proposed by Petersen \textit{et al.}~\cite{petersen2015}. The process for conducting this study consists of three phases: planning, conducting and reporting the mapping. To support the selection of relevant studies, we have adopted the recommendations proposed by Zhang \textit{et al.}~\cite{zhanghe2011} to establish a more precise and efficient search protocol. Those phases are explained in more details in the following subsections.

\subsection{Planning the Mapping}
\label{sec:planning-the-mapping}

The protocol of this research was developed and validated by the five authors. Our protocol is composed by: (i) definition of research questions, (ii) search strategy, (iii) selection of studies, (iv) data extraction and (v) data synthesis and aggregation strategy. In the following subsections each of the phases will be described.

\subsubsection{Research Questions}

This systematic mapping aims at providing a glimpse on how solutions based on self-adaption techniques are being applied in the domain of microservice-based systems, i.e., we want to understand and analyse how microservices are using the self-adaptability concepts. Based on this, ten primary research questions presented in Table \ref{tab:tableresearch} were formulated to guide the research methodology.

The research questions formulated address the adoption of self-adaptive techniques in microservices-based architectures and provide specific details on the subject when applied in this new context, in addition to showing how the current scenario on the research topic is, addressing how the studies has contributed for the development of new solutions.

\subsubsection{Search Strategy}

The research strategy is defined according to the process defined by~\cite{zhanghe2011}, which suggest the resolution of the following questions: \textit{Which} approach to use in the search process; \textit{Where} (locations or databases) to search and which part of the article should be searched; \textit{What} (subject, type of evidence) should be searched for and what queries (search strings) are inputted into the search engines; \textit{When} the search is performed and the period to be searched for.

\noindent\textbf{\textit{Which approach(es)?}} We used  automatic search involving electronic resources, such as digital libraries and indexing systems, to search for relevant articles~\cite{kitchenham2007}. In addition, the snowballing process, backwards and forwards, was used in the selected papers for reading~\cite{kitchenham2007}.

\noindent\textbf{\textit{Where to search?}} The automatic search was performed in three digital libraries (IEEE Xplore, ACM Digital Library, ScienceDirect) and two indexing mechanisms (Scopus and Springer Link). According to~\cite{BRERETON2007571}, those are the recommended data sources of papers about software.

\noindent\textbf{\textit{What to search?}} Based on the objectives of this systematic mapping study, a set of keywords were analyzed, selected and organized into groups, as shown in Table \ref{tab:tablesearchstring}, then the search string was formulated. The first group refers to words associated with the microservice context, while the second group involves words related to the domain of self-adaptive systems.

\noindent\textbf{\textit{When and which period to search?}} The research was conducted between February 2020 and May 2020.

\subsubsection{Study Selection}

We adopted one inclusion criteria and five exclusion criteria to select relevant articles to answer the survey questions. The inclusion criteria is: papers that explicitly discuss a self-adaptive approach or technique applied in the context of microservices. The exclusion criteria are: (i) studies published only as a (short) abstract; (ii) studies not written in English; (iii) studies to which it was not possible to have access; (iv) duplicate studies; (v) studies that only mention adaptation mechanisms and techniques as a general introductory term and/or are not focused on applying them to the microservice architectural style. Studies that do not meet any exclusion criteria, but agree with the inclusion criterion, are considered relevant.

\subsubsection{Data Extraction}
\label{sec:data-extraction}

Firstly, during the mapping and data extraction stage, some  authors were sub-divided into pairs and carefully examined each of the studies belonging to the final set, randomly assigned to each pair. 
Each pair of reviewers individually analyzed their study group. The divergences found during the process were resolved through discussions among all authors.

\subsection{Conducting the Mapping}
\label{sec:conducting-the-mapping}

\begin{table}[]
\caption{Search results by electronic database at each phase of selection}
\label{tab:tablebydatabase}
\begin{tabular}{lllll}
\hline
\textbf{Database} & \textbf{P1} & \textbf{P2} & \textbf{P3} & \textbf{P4 (\#Snowballing)} \\ \hline
IEEE & 46 & 41 & 13 & 6 (2) \\
ACM Digital Library & 226 & 29 & 9 & 4 \\
SpringerLink & 405 & 20 & 3 & 4 \\
ScienceDirect & 21 & 18 & 4 & 1 \\
Scopus & 117 & 42 & 8 & 4 \\
\hline
\textbf{Total} & 815 & 150 & 37 & 21 \\ 
\hline
\end{tabular}
\end{table}

At this stage, the mapping protocol defined in the planning phase was executed. In general, studies were searched, selected and analyzed by the researchers of this work in order to extract and synthesize the relevant data for answering research questions. For the search and selection of articles, a four-phase process was applied. Table \ref{tab:tablebydatabase} summarizes the number of studies obtained in each database, for each phase, and the total number of studies included in our mapping study.

In the first phase (P1), an automated search was performed by executing the search string in the electronic databases defined in the protocol. As the term ``microservice'' is considered relatively new, no restriction on the year of publication was defined, thus allowing a wide range of articles to be returned. At the end of this phase, a set of 815 articles was returned.

In the second phase (P2), manual filtering was performed on the bases for which it was not possible to search studies automatically. Studies that did not have at least one term from each group of the search string were removed from the set of studies for the next phase. At the end of filtering, a total of 150 works were considered studies suitable for the next phase.

In the third phase (P3), for each selected study it was read the title, abstract and keywords to filter the studies based on the selection criteria (inclusion/exclusion). This phase was initially conducted by two of the authors in order to minimize the effect of any bias or misinterpretation. A meeting was then held among the researchers involved to compare the results and resolve existing conflicts, obtaining a consensual preliminary selection. When there was divergence, a third researcher was asked to define the final result. At the end of this phase, after removal of duplicate papers, 37 papers were evaluated as potentially relevant.

\begin{table*}[]
\centering
\caption{Selected studies}
\label{tab:tableselectedstudies}
\begin{tabular}{p{0.5cm} p{0.6cm} p{0.6cm} p{1.4cm} p{2.2cm} p{2cm} p{2.2cm} p{1.7cm} p{2.5cm}}
\hline
{\footnotesize \textbf{ID}} & {\footnotesize \textbf{Ref.}} & {\footnotesize \textbf{Year}} & {\footnotesize \textbf{Publication Type}} & {\footnotesize \textbf{Self-* Property}} & {\footnotesize \textbf{Mechanism}} & {\footnotesize \textbf{Research Type}} & {\footnotesize \textbf{Contribution Type}} & {\footnotesize \textbf{Research Method}} \\ \hline

{\footnotesize S1}  & {\footnotesize \cite{s1journal}} & {\footnotesize 2019} & {\footnotesize journal} & {\footnotesize self-optimizing} & {\footnotesize MAPE-K control loop} & {\footnotesize solution proposal} & {\footnotesize tool} & {\footnotesize case study} \\ 

{\footnotesize S2}  & {\footnotesize \cite{s2conference}} & {\footnotesize 2018} & {\footnotesize conference} & {\footnotesize not specified} & {\footnotesize feedback loop} & {\footnotesize solution proposal} & {\footnotesize model, tool, method} & {\footnotesize -} \\

{\footnotesize S3}  & {\footnotesize \cite{s3journal}} & {\footnotesize 2017} & {\footnotesize journal} & {\footnotesize self-management, self-healing} & {\footnotesize undefined} & {\footnotesize experience report} & {\footnotesize evaluation, model, lessons learned} & {\footnotesize experiment} \\

{\footnotesize S4}  & {\footnotesize \cite{s4conference}} & {\footnotesize 2019} & {\footnotesize conference} & {\footnotesize self-healing} & {\footnotesize MAPE-K control loop} & {\footnotesize solution proposal} & {\footnotesize model, method} & {\footnotesize case study} \\

{\footnotesize S5}  & {\footnotesize \cite{s5conference}} & {\footnotesize 2016} & {\footnotesize conference} & {\footnotesize not specified} & {\footnotesize No mechanism applied} & {\footnotesize philosophical paper} & {\footnotesize review} & {\footnotesize -} \\

{\footnotesize S6}  & {\footnotesize \cite{s6conference}} & {\footnotesize 2017} & {\footnotesize conference} & {\footnotesize not specified} & {\footnotesize feedback loop} & {\footnotesize solution proposal} & {\footnotesize model, evaluation, method} & {\footnotesize -} \\

{\footnotesize S7}  & {\footnotesize \cite{s7journal}} & {\footnotesize 2019} & {\footnotesize journal} & {\footnotesize not specified} & {\footnotesize No mechanism applied} & {\footnotesize philosophical paper} & {\footnotesize review, lessons learned} & {\footnotesize -} \\

{\footnotesize S8}  & {\footnotesize \cite{s8conference}} & {\footnotesize 2018} & {\footnotesize conference} & {\footnotesize not specified} & {\footnotesize undefined} & {\footnotesize solution proposal} & {\footnotesize model, tool} & {\footnotesize experiment $<$ 50\%} \\

{\footnotesize S9}  & {\footnotesize \cite{s9conference}} & {\footnotesize 2018} & {\footnotesize conference} & {\footnotesize not specified} & {\footnotesize MAPE-K control loop} & {\footnotesize solution proposal} & {\footnotesize review, method} & {\footnotesize -} \\

{\footnotesize S10}  & {\footnotesize \cite{s10conference}} & {\footnotesize 2016} & {\footnotesize conference} & {\footnotesize not specified} & {\footnotesize MAPE control loop} & {\footnotesize solution proposal} & {\footnotesize tool} & {\footnotesize experiment $<$ 50\%} \\

{\footnotesize S11}  & {\footnotesize \cite{s11conference}} & {\footnotesize 2016} & {\footnotesize conference} & {\footnotesize not specified} & {\footnotesize MAPE-K control loop} & {\footnotesize philosophical paper} & {\footnotesize review} & {\footnotesize -} \\

{\footnotesize S12}  & {\footnotesize \cite{s12workshop}} & {\footnotesize 2015} & {\footnotesize workshop} & {\footnotesize self-management, self-healing} & {\footnotesize undefined} & {\footnotesize solution proposal} & {\footnotesize model} & {\footnotesize -} \\

{\footnotesize S13}  & {\footnotesize \cite{s13conference}} & {\footnotesize 2018} & {\footnotesize conference} & {\footnotesize not specified} & {\footnotesize No mechanism applied} & {\footnotesize philosophical paper} & {\footnotesize model, review} & {\footnotesize -} \\

{\footnotesize S14}  & {\footnotesize \cite{s14conference}} & {\footnotesize 2018} & {\footnotesize conference} & {\footnotesize not specified} & {\footnotesize MAPE control loop} & {\footnotesize solution proposal} & {\footnotesize model} & {\footnotesize case study $<$ 50\%} \\

{\footnotesize S15}  & {\footnotesize \cite{s15conference}} & {\footnotesize 2019} & {\footnotesize conference} & {\footnotesize not specified} & {\footnotesize MAPE control loop} & {\footnotesize solution proposal} & {\footnotesize model, language, tool, framework} & {\footnotesize case study $<$ 50\%} \\

{\footnotesize S16}  & {\footnotesize \cite{s16conference}} & {\footnotesize 2018} & {\footnotesize conference} & {\footnotesize not specified} & {\footnotesize MAPE control loop} & {\footnotesize solution proposal} & {\footnotesize model, framework} & {\footnotesize -} \\

{\footnotesize S17}  & {\footnotesize \cite{s17conference}} & {\footnotesize 2019} & {\footnotesize conference} & {\footnotesize self-healing} & {\footnotesize undefined} & {\footnotesize solution proposal} & {\footnotesize review, model} & {\footnotesize -} \\

{\footnotesize S18}  & {\footnotesize \cite{s18conference}} & {\footnotesize 2017} & {\footnotesize conference} & {\footnotesize self-healing} & {\footnotesize self-healing loop} & {\footnotesize solution proposal} & {\footnotesize framework} & {\footnotesize -} \\

{\footnotesize S19}  & {\footnotesize \cite{s19conference}} & {\footnotesize 2020} & {\footnotesize conference} & {\footnotesize self-management, self-optimizing} & {\footnotesize undefined} & {\footnotesize solution proposal} & {\footnotesize model} & {\footnotesize -} \\

{\footnotesize S20}  & {\footnotesize \cite{s20journal}} & {\footnotesize 2020} & {\footnotesize journal} & {\footnotesize not specified} & {\footnotesize MAPE-K control loop} & {\footnotesize solution proposal} & {\footnotesize model, metrics, evaluation} & {\footnotesize experiment $<$ 50\%} \\

{\footnotesize S21}  & {\footnotesize \cite{s21journal}} & {\footnotesize 2017} & {\footnotesize journal} & {\footnotesize not specified} & {\footnotesize MAPE-K control loop} & {\footnotesize solution proposal} & {\footnotesize model} & {\footnotesize -} \\

\hline
\end{tabular}
\end{table*}

In the fourth phase (P4), the filtered studies were entirely read. This phase was initially conducted by three researchers, allocated in pairs. Agreement meetings were held to compare results and resolve conflicts. In case of divergences, a fourth researcher decided whether or not to include the study in the final set. This process resulted in 19 papers considered relevant. Finally, we applied the forward and backward snowballing technique considering the selected papers. At the end, we obtained two more new papers, resulting a total of 21 relevant studies belonging to our final set. 
Table \ref{tab:tableselectedstudies} gives an overview of the selected studies including some information that, due to space limitation, we could not discuss in this paper, such as year and type of publication.

\subsection{Reporting the Results}

In this section we present and discuss the results of the research questions that drive this work.\newline

\noindent\textbf{\textit{RQ1: How has self-adaptation been applied in the context of microservices?}} 

Adalberto \textit{et al.} [S1] proposed an adaptation mechanism, called REMaP (RuntimE Microservices Placement), which improves the management of applications based on microservices through the use of historical and runtime data as the basis for autonomic management tasks. One of the main challenges that the authors addressed is to find a better allocation of microservices among the available servers aiming a performance gain. For this, the tool uses a MAPE-K based adaptation manager to alter the deployment of microservices at runtime autonomously and groups and allocates microservices with high affinity on the same physical server. In [S2], Kehrer and Blochinger proposed a concept for self-adaptation of cloud applications based on container virtualization, according to the architectural style of microservices, and then presented a model-based approach that assists software developers on building services. Based on the operational models specified by the developers, the necessary mechanisms for self-adaptation are generated automatically.

Toffetti \textit{et al.} [S3] discussed the main characteristics of cloud native applications, proposing a new architecture that allows scalable and resilient self-management applications in the cloud and reported the experiences of porting a legacy application to the cloud, applying cloud native principles. In [S4], Magableh and Almiani proposed a self-healing microservice architecture, by following the MAPE-K model.

In [S5], Kizilov \textit{et al.} investigated the use of Domain Objects as an alternative to support the adaptation of component-based services and capture peculiarities of the microservice architecture. In this way, the authors performed a general explanation of Domain Objects and discussed ways to extend an adaptive framework to be used with microservices. In [S6], Nguyen and Nahrstedt proposed the design and implementation of MONAD, a self-adaptive infrastructure based on microservices to execute heterogeneous scientific workflow. To deal with various constraints, the authors designed a resource adaptation approach based on feedback control loop. The approach has layers of monitoring and adaptation in order to incorporate performance guarantees to the adaptation objective and seeks to find resource allocation strategies that meet resource budget constraints.

Mendonça \textit{et al.} [S7]  explore the relationship and interaction between self-adaptive systems and microservices in order to identify the main challenges for the development of microservice applications as self-adaptive systems. The main challenges addressed are related to (i) the determination of monitoring and adaptation mechanisms to deal with the diversity of microservices' quality attributes; (ii) identify and resolve conflicts of individual and global requirements between microservices and the system; and (iii) how to manage and apply selected control loop mechanisms in an already determined and existing infrastructure management environment. In [S8], Xianghui \textit{et al.} proposed a model of evolution based on distributed knowledge (DKEM) to promote the healthy evolution of an ecosystem of services, including distributed knowledge. That model depends on local and global evolution patterns to automatically evolve the ecosystem when a service-based process is performed in a self-adaptive manner.

In [S9], Khazaei \textit{et al.} proposed and supported the idea of providing adaptation for cloud applications as a service. Another aspect of the proposed idea is to move from self-adaptation, in which the software itself adapts to the new conditions for ADaptation-as-a-Service (ADaaS), in which the environment can protect, optimize, configure and cure the software system following the principles of the MAPE-K model.  Florio and Di Nitto presented in [S10] a methodology and tool for applying autonomic computing to applications based on the microservice architecture pattern. They proposed a framework, Gru, which has self-adaptive capabilities, such as automatic container sizing, as well as self-healing features. The MAPE control loop mechanism is implemented internally to the framework components, called Gru-Agents, which implement the MAPE classic autonomous loop that is triggered periodically, with a time interval defined by the user in the configuration of the agents. Gru is able to manage a microservice application deployed in Docker containers and has been validated through an initial test to show its capabilities and potential. 

In [S11], Hassan and Bahsoon conducted a review and reflection on the definition, properties and techniques of microservice modeling. In addition, they discussed decisions related to architectural decision-making regarding the granularity of microservices in systems, and how that it can be decided based on non-functional requirements and how self-adaptability can be an aid factor in decisions. In [S12], Toffetti \textit{et al.} proposed a new architecture that allows scalable and resilient self-managing microservices to function in the cloud. The proposed new architecture has monitoring, health management and dimensioning features that are naturally adapted to the managed application and its dynamic nature.

In [S13], Mendonça \textit{et al.} discussed and analyzed architecture-based self-adaptation services and frameworks that have been proposed in recent years from two design perspectives: generality and reusability. Through this, they identified several patterns to add self-adaptation capabilities to existing systems. In [S14], Peini \textit{et al.} proposed a reference architecture for self-adaptability in microservice-based systems with multi-layer controlled self-adaptation skills, including infrastructure-controlled layer and application-controlled layer. The authors applied the MAPE control loop to both infrastructure and  application levels. The authors also presented results of experiments conducted to validate the solution.

In [S15], Zhang \textit{et al.} proposed a self-adaptive approach that works at different levels to enrich the self-adapting capabilities of microservice-based architecture and technologies. The authors present a tool called AdaptiveK8s, which is implemented as an extension of Kubernetes and that, through the MAPE control loop, enables the ability to make multi-level adaptations of the application. The tool monitors data related to microservice requirements and updates, in addition to providing the ability to manage aspects of scalability. In [S16], Zhang \textit{et al.} analyzed the special self-adaptive requirements for microservice-based systems and proposed a self-adaptive microservice framework, called MSSAF, based on a reference architecture that uses the MAPE control loop. The self-adaptation mechanism consists of several components that provide the self-adaptation monitoring, detection, decision and action capabilities.

In [S17], Wang presented a self-healing solution for microservice-based systems. The initial focus of the research was to activate self-healing executions from the moment when updates were made in the system. The architectural proposal presented has the ability to react autonomously to changes, including failures, scale variations and evolution.  Li \textit{et al.} propose in [S18] an improved service framework with a self-healing feature, adding a set of customizable and flexible plugins to the common cloud orchestration function. The presented framework allows users to define the failure detection approach and the  list of recovery actions to be attempted when failures are detected. The microservice model is used to improve delivery efficiency in addition to helping fault analysis for cloud platform administrators and application owners.

In [S19], Sanctis \textit{et al.} proposed a new self-adaptive architecture for microservice based IoT systems. The architecture deals with proactive adaptation using machine learning (ML) techniques and reactive adaptation, exploring the dynamic composition of microservices. In addition, the architecture presented takes into account adaptation concerns that may arise at different levels. In [S20], Magableh and Almiani presented a set of base tests for distributed and self-adaptive architectures for microservices that can be used by other researchers to experiment with various self-adaptation techniques. In addition, they introduced adaptation agents that implement the Markov decision process that are able to collect observations about the environment and can carry out adaptation actions.

Finally, in [S21], Baylov and Dimov proposed a reference architecture for microservice systems based on autonomic computing that provides a way to transform microservice instances by adding an autonomous manager to a part of the service and building an adaptation record. This approach makes it possible to transform existing production systems into self-adaptive systems in several iterations and with a low risk. \newline

\noindent\textbf{\textit{RQ2: What types of research and contribution have been presented?}}

The research type represents the approach used in the studies to be analyzed. In this work, we  adopt the classification scheme proposed by Wieringa \textit{et al.}~\cite{wieringa2006}, which divides the research types into six categories: evaluation, validation, solution proposal, philosophical paper, opinion paper and experience reports. Regarding the contribution type, we use the classification method suggested by Shahrokni and Feldt~\cite{shahrokni2013}, which is organized into seven categories: review, model, metrics, evaluation, framework, tool, method. In addition to that classification of contributions, we adopt two other types: lessons learned and language. We consider that a work can be classified as having more than one contribution type.

\begin{figure}[htbp]
\centerline{\includegraphics[scale=0.50]{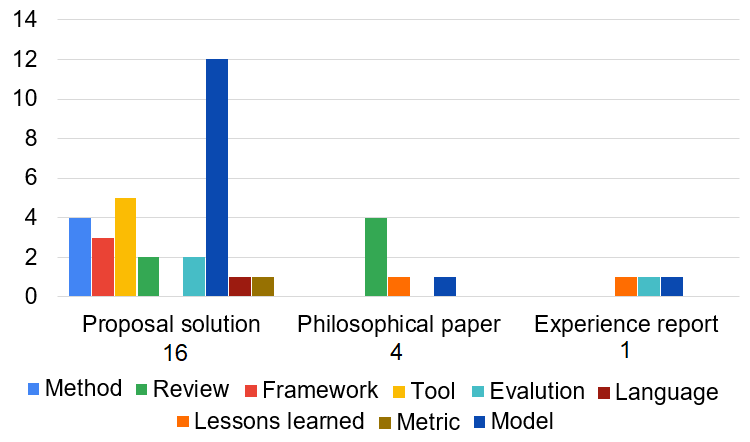}}
\caption{Research type vs contribution type}
\label{fig:pesquisa_vs_contribuicao}
\end{figure}

Figure \ref{fig:pesquisa_vs_contribuicao} shows that most of the studies classified from the research type view were solution proposals with 16 studies (76.19\%), followed by philosophical papers with 4  studies (19.05\%) and only 1 experience report (4.76\%). Regarding the contribution type view, still according to Figure \ref{fig:pesquisa_vs_contribuicao}, it is possible to observe that the 3 most expressive types of contribution were models, with 14 papers (66.67\%), then 6 studies presented reviews (28.57\%), followed by 5 studies that proposed tools (23.81\%). Finally, 3 studies (14.29\%) did not apply any type of control loop mechanism.

Taking into account only studies of solution proposal, we noticed that most of them have models as contribution (75\%). In contrast, only 5 studies proposed some type of tool. These results may indicate that research in this area is still at an architectural model stage and that it should reach a higher maturity level, in terms of tools and frameworks in the coming years, through the validation of the proposed models. \newline

\noindent\textbf{\textit{RQ3: Which phase of a self-adaptation control loop is the focus of the studies?}} 

This research question aims at identifying where the contribution of each study is taking into account the phases of the control loop. Table \ref{tab:table_evident_phases} shows the results. 
Note that nine studies (42.86\%) did not focus on a specific phase, which means that either the solution traverses all phases or the contribution of the work is not within the feedback control loop, which is used only as a mechanism to implement the self-adaptation.

Also according to Table \ref{tab:table_evident_phases}, six studies (28.57\%) focused on the Monitoring stage. There was only one study that focused on the combination of the Monitor and Analysis  phases; one study whose contribution touches the Monitoring, Analysis and Planning phases; and one study that addressed the Analysis and Planning stage. It is still possible to see in Table \ref{tab:table_evident_phases}, that no study focused on the stages of analysis or planning in isolation. Finally, three studies (14.29\%) did not apply any type of control loop. These studies are philosophical papers that did not use any phase and did not give further details on the topic.

Note that, according to Table~\ref{tab:tableselectedstudies}, the MAPE control loop and its variation with a knowledge base (MAPE-K) was the most used one, being applied by ten studies (47.6\%). Two studies (9.52\%) used a generic feedback loop and only one study (4.76\%) used a so-called self-healing control loop, which is specific for dealing with the self-* property of the same name. In total, five studies (23.81\%) did not explicitly define which control loop mechanism was used.

Among the three phases that received the most attention, only the Monitoring phase was evidenced in isolation, while the other phases work in conjunction. Due to the distributed nature of this architectural style, some approaches proposed Monitoring at different levels: infrastructure, through the extraction of metrics from the use of containers ([S14], [S10]); platform-specific devices, such as IoT devices ([S19]); and application, by monitoring metrics related to functional and non-functional requirements ([S19], [S14]). \newline

\begin{table}[]
\caption{Distribution of studies according to the most evident phase of the feedback control loop}
\label{tab:table_evident_phases}
\begin{tabular}{p{3.0cm} p{0.3cm} p{3.0cm}}
\hline
\textbf{Phase} & \textbf{\#} & \textbf{Studies} \\ \hline
No evident phase & 9 & S2, S4, S9, S10, S11, S15, S16, S20, S21 \\
Monitor & 6 & S3, S6, S8, S12, S17, S18 \\
Monitor and Analyze & 1 & S14 \\
Monitor, Analyze and Plan & 1 & S19 \\
Analyze and Plan & 1 & S1 \\
Not applied & 3 & S5, S7, S13 \\
\hline
\end{tabular}
\end{table}

\noindent\textbf{\textit{RQ4: What self-* properties have been addressed?}}

As depicted by Table \ref{tab:table_self_properties}, only three self-* properties have been explicitly addressed by the studies: self-healing (five studies - 23.81\%), self-management (three studies - 14.29\%) and self-optimizing (two studies - 9.52\%). 

In the context of microservices, the studies addressed the self-healing property for anomaly detection [S3], for resilient scalability [S4], reallocation of microservices in different deployment environments [S12], ability to self-update versions of microservices without impacting the availability of the application [S17], addition of customizable and flexible plugins to the orchestration [S18]. The studies [S3] and [S12] deal with self-management according to load variations and by managing components when necessary, such as creating new components or destroying them; [S19] has the capacity of self-managing microservice applications to multilevel IoT systems based on machine learning techniques.
Finally, the studies [S1] and [S19] deal with self-optimizing property. While [S1] focused on optimizing the allocation of microservices, [S19] proposed to optimize the performance of the system through the monitoring of various QoS qualities parameters, such as traffic and response time.

Note that some studies address more than one property, such as [S3], [S12] and [S19]. On the other hand, 14 studies (66.67\%) did not specify which self-* property was the focus of the adaptation proposed. \newline 

\begin{table}[]
\caption{Self-* properties addressed by the studies}
\label{tab:table_self_properties}
\begin{tabular}{p{2cm} p{0.3cm} p{5.0cm}}
\hline
\textbf{Property} & \textbf{\#} & \textbf{Studies} \\ \hline
Self-healing & 5 & S3, S4, S12, S17, S18 \\
Self-management & 3 & S3, S12, S19 \\
Self-optimizing & 2 & S1, S19 \\
Not specified & 14 & S2, S5, S6, S7, S8, S9, S10, S11, S13, S14, S15, S16, S20, S21 \\
\hline
\end{tabular}
\end{table}

\noindent\textbf{\textit{RQ5: What self-adaptation strategies have been used?} }

Regarding the adaptation strategy, according to Table \ref{tab:table_adaptation_strategy}, the vast majority of studies used reactive adaptation (80.95\%), which indicates that the proposed solutions only solve problems at the moment they occur. On the other hand, only one study (4.76\%) proposed a proactive adaptive solution and another one (4.76\%) used a hybrid solution. Although proactive adaptation is obviously preferable as it avoids interruptions in the user's workflow with the system, reactive adaptation is much easier to achieve and, consequently, is preferred in most approaches to self-adaptive systems~\cite{krupitzer2015}. Finally, for two studies (9.52\%) it was not possible to identify which strategy was used. \newline

\begin{table}[]
\caption{Adaptation strategies adopted by the studies}
\label{tab:table_adaptation_strategy}
\begin{tabular}{p{3.0cm} p{0.3cm} p{3.0cm}}
\hline
\textbf{Adaptation} & \textbf{\#} & \textbf{Studies} \\ \hline
Reactive & 17 & S1, S2, S3, S4, S5, S8, S9, S10, S12, S13, S14, S15, S16, S17, S18, S20, S21 \\
Proactive & 1 & S6 \\
Hybrid & 1 & S19 \\
Unidentified or Unused & 2 & S7, S11 \\
\hline
\end{tabular}
\end{table}

\noindent\textbf{\textit{RQ6: What quality requirements have been addressed by the approaches?}}

Table \ref{tab:table_concerns_addressed} shows the results of the quality attributes that were addressed in the selected studies. We can see that the attributes of performance (12 studies - 57.14\%), scalability (10 studies - 47,62\%) and resilience (8 studies - 38.10\%) are the main concern of most of the studies. This fact is not surprising, since those attributes are benefits sought when developing microservice-based systems. 

Two studies addressed security problems, and the same number focused on reliability issues. Reusability, portability and flexibility were the focus of only one study each. Finally, four studies did not specify the quality attribute addressed. Note that is very common that the same study approaches more than one quality attribute. For example, [S2] addressed performance, reusability and portability, while [S9] dealt with portability, scalability, security and reliability.  \newline

\begin{table}[]
\caption{Quality requirements addressed by the studies}
\label{tab:table_concerns_addressed}
\begin{tabular}{p{2.0cm} p{0.3cm} p{5.0cm}}
\hline
\textbf{Requirement} & \textbf{\#} & \textbf{Studies} \\ \hline
Performance & 12 & S1, S2, S4, S6, S9, S10, S13, S15, S16, S18, S19, S20 \\
Scalability & 10  & S3, S4, S6, S9, S10, S12, S13, S14, S16, S17 \\
Resilience & 8  & S3, S4, S8, S12, S14, S17, S18, S19 \\
Security & 2 & S9, S13 \\
Reliability & 2 & S9, S18 \\
Reusability & 1 & S2 \\
Portability & 1 & S2 \\
Flexibility & 1 & S6 \\
Not specified & 4 & S5, S7, S11, S21 \\
\hline
\end{tabular}
\end{table}

\noindent\textbf{\textit{RQ7: In which microservice architecture layer were the adaptations applied?}}

Table \ref{tab:table_layer_applied} informs the architecture layer in which self-adaptation was applied in the selected studies. It can be seen that the vast majority of studies (10 studies - 47.62\%) proposed to carry out self-adaptations in terms of the system's infrastructure, which is strictly linked to the microservices instances in execution, and regards containers and their orchestration tools. This is in accordance with the answer of RQ6, which showed that performance, scalability and resilience, attributes that are strongly linked to the infrastructure layer, are the most addressed ones.

In contrast,  no study has presented self-adaptation strategies only at the application level, which concerns the functionality of microservices, but five studies (23.81\%) presented multi-layer self-adaptation approaches. This shows that it is very hard to decouple the microservices from their underlying infrastructure such that an adaptation, even in the application layer, always reflects on it counterpart layer. Two studies (9.52\%) did not make it clear in which application layer the self-adaptations were intended, therefore they are included in the ``Not specified'' category. At last, four studies (19.05\%) in the ``Not applicable'' category are philosophical papers and do not discuss in which layer of microservices the adaptations are applied. \newline

\begin{table}[]
\caption{Microservice layers in which the adaptation occurred}
\label{tab:table_layer_applied}
\begin{tabular}{p{2.0cm} p{0.3cm} p{5.0cm}}
\hline
\textbf{Layer} & \textbf{\#} & \textbf{Studies} \\ \hline
Infrastructure & 10 & S1, S2, S3, S4, S6, S10, S12, S18, S20, S21 \\
Multilayer & 5 & S14, S15, S16, S17, S19 \\
Application & 0 & - \\
Not specified & 2 & S8, S9 \\
Not applicable & 4 & S5, S7, S11, S13 \\
\hline
\end{tabular}
\end{table}

\noindent\textbf{\textit{RQ8: What self-adaptation control logic has been addressed?}}

Table \ref{tab:table_control_logic_pattern} provides an overview of which type of self-adaptation control logic have been addressed by the studies. The centralized control logic is the most used one with 38.10\% (8 studies) of the total. This may be explained by the fact that it is easier to extract monitoring data through the container orchestration and infrastructure provisioning tools, as well as using the APIs of these tools to perform the necessary adaptations.

Decentralized control logic, in which agents are active in different parts of the system, were addressed  by 3 studies (14.29\%). Implementing such strategies usually requires a sidecar at each microservice instance to perform, for example, data collection from logging and health check, in order to change the behavior of the running instance.

Hybrid control logic, which combines the two previous ones, appear in 2 studies (9.52\%). It was not possible to see in 4 studies (19.05\%) which control logic had been addressed, thus they are included in the ``Not specified'' category. The lack of information regarding the way in which the logic of adaptation has been applied may justify this found. At last, four studies (19.05\%) in the ``Not applicable'' category are philosophical papers and do not discuss which control logic had been addressed. \newline

\begin{table}[]
\caption{Adaptation control logic used by studies}
\label{tab:table_control_logic_pattern}
\begin{tabular}{p{2.0cm} p{0.3cm} p{4.5cm}}
\hline
\textbf{Pattern} & \textbf{\#} & \textbf{Studies} \\ \hline
Centralized & 8 & S1, S14, S15, S16, S17, S18, S19, S21 \\
Decentralized & 3 & S2, S6, S10 \\
Hybrid & 2 & S4, S20 \\
Not specified & 4 & S3, S8, S9, S12 \\
Not applicable & 4 & S5, S7, S11, S13 \\
\hline
\end{tabular}
\end{table}

\noindent\textbf{\textit{RQ9: What technologies have been addressed?}}

Among the 21 analyzed studies, 16 of them (76.19\%) mentioned specific technologies in the contributions, as depicted by Table \ref{tab:table_overview_techonologies}. We classified the technologies in 15 different categories. It can be seen that the majority of the studies addresses Container technologies (14 studies - 66.6\%), followed by Storage technologies (9 studies - 42.85\%). This reflects the finding of RQ7 that reported that infrastructure is the layer most addressed layer.

Another highlight is that technologies for monitoring, collecting and analyzing application metrics were addressed by 38.09\% of the studies. This is also in tandem with the RQ3, which indicates that the Monitoring phase of the feedback loop is the most addressed one.

\begin{table*}[ht]
\caption{Overview of technologies covered in the studies}
\label{tab:table_overview_techonologies}
\begin{tabular}{p{4.0cm} p{5.0cm} p{0.3cm} p{5.0cm}}
\hline
\textbf{Category} & \textbf{Technologies} & \textbf{\#} & \textbf{Studies} \\ \hline
Container & Docker, Kubernetes, Docker Swarm, Apache Mesos, ContainerPilot, Fleet & 14 & S1, S2, S3, S4, S6, S7, S9, S10, S12, S13, S14, S15, S19, S20 \\
Storage & ETCD, InfluxDB, Zookeeper, Apache Derby, ELK & 9 & S1, S3, S6, S9, S10, S12, S14, S19, S21 \\
Metrics & cAdvisor, Beats, Prometheus, Collectd, Zipkin, Heapster, Consul Watches, Elasticsearch, Grafana, ELK & 8 & S1, S9, S14, S20, S4, S3, S9, S19 \\
Service Mesh & Istio, Consul, Linkerd & 4 & S2, S7, S13, S19 \\
Devops Tools & Spinnaker, Travis, Jenkins & 3 & S1, S7, S9 \\
Machine Learning & Tensorflow, SciPy, Kera (Python) & 3 & S6, S20, S19 \\
Cloud Infrastructure & AWS, Harbor, Openstack, Heat, Senlin, Mistral & 3 & S3, S14, S18 \\
Distributed Processing & RabbitMQ, Kafka & 2 & S6, S19 \\
Web Framework & Apache Freemaker, Jersey & 2 & S2, S21 \\
Business Rules Management & Drools & 1 & S14 \\
Logging & Log-courier & 1 & S3 \\
Modeling Tool & EMF Framework & 1 & S1 \\
Operations Management & vROPS & 1 & S18 \\
Specification Language & TOSCA & 1 & S2 \\
Test & Apache JMeter & 1 & S10 \\
Not Specified & - & 5 & S5, S8, S11, S16, S17 \\
\hline
\end{tabular}
\end{table*}

Furthermore, it is interesting to see that, even being a relative recent technology for intercommunication of microservices, service mesh comes in fourth place, being addressed by 4 studies (19.04\%). Following, with 3 studies each, come the categories DevOps and Machine Learning tools, which indicates that those concepts are gaining attention in the approaches. Complete the group with three studies the category Cloud Infrastructure.

Then, a group of eight technologies that have been addressed by only two or one study appears. Finally, five studies (23.80\%) did not address or discuss technologies in their contributions, being two philosophical papers and three solution proposals. \newline

\noindent\textbf{\textit{RQ10: Has an empirical evaluation been conducted?}}

\begin{figure}[htbp]
\centerline{\includegraphics[scale=0.25]{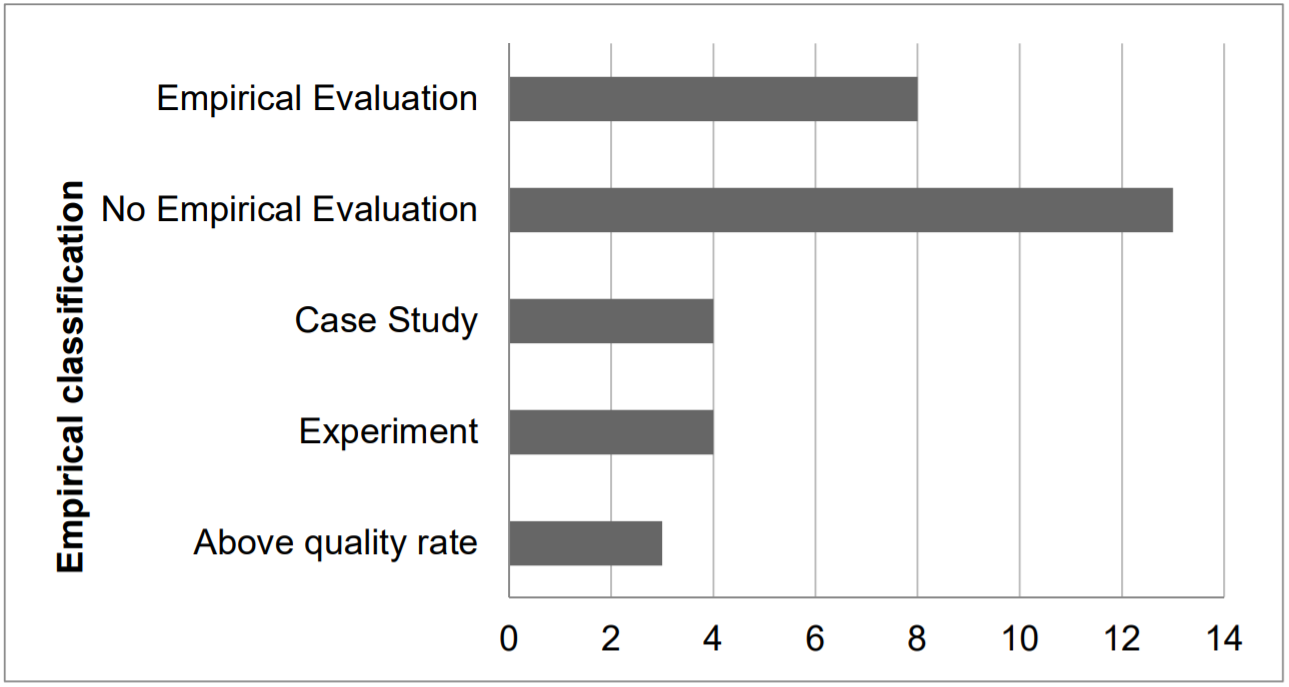}}
\caption{Empirical evaluation of the studies}
\label{fig:avaliacao_estudos}
\end{figure}

In order to answer RQ10, we carried out a process of quality assessment and classification of the empirical methods used to validate the contribution. Likewise for the evaluation methods, we classified the studies according to the categories defined by Wohlin \textit{et al.} \cite{wohlin2012}: experiment, case study and survey. For the quality assessment process, we have defined \textit{checklists} with questions for each empirical method based on literature quality assessment criteria \cite{kitchenham2007}. Each question is evaluated as ``Yes'' (indicating that data for the specific question are clearly available), ``Partly'' (data are vaguely available) or ``No'' (data are not available) with a corresponding score of 1, 0.5 or 0, respectively. Based on the quality criteria established by Kitchenham \textit{et al.} \cite{kitchenham2007}, this scale is necessary because a simple answer Yes/No can be misleading.

Figure \ref{fig:avaliacao_estudos} shows the number of studies that evaluated their proposals with empirical evaluation methods. From the 21 studies selected, only 8 (38.09\%) performed empirical evaluation, while 13 (61.90\%) did not perform any of the three types of empirical evaluation. Despite this, we obtained studies that demonstrated the evaluation of their proposal through use example ([S16] and [S11]) and 1 study that performed a proof of concept [S9]. Thus, following the quality criteria adopted in this work, the cited studies did not conduct an empirical evaluation, be it a case study, experiment or survey. \newline

\noindent\textbf{\textit{RQ10.1: What strategy was used to validate the research?}}

Still according to Figure \ref{fig:avaliacao_estudos}, we can see that, from the 8 studies that did in fact perform an empirical evaluation, 4 studies performed a case study (50\%), while the other 4 (50\%) performed an experiment. There were no studies that carried out a survey. Regarding this set of work, only 37.5\% (3 studies) obtained a quality rate above 50\%, according to our quality evaluation process, being 2 case studies, [S1] and [S4], and 1 experiment [S3]. Because most studies are solution proposals that require at least some type of validation, it is evident that more studies are needed to prove the real effectiveness of their solutions, whether they are models, tools, methods or any type of contribution and artifact that can be evaluated empirically.

\section{Research Directions}
\label{sec:discussion}

After drawing the general picture of how self-adaptation mechanisms are currently being  applied to microservice-based systems through the research questions answered by the systematic literature review, it is possible to identify gaps that open new research opportunities described as follows.

\subsection{Resilience, security and reliability in microservice composition}

As demonstrated by Table~\ref{tab:table_layer_applied}, only five studies proposed solutions that address the microservice application level through a multi-layer approach. At the application layer, the system is concerned with the composition of microservices, using either orchestration or choreography, to realize business rules. Although some studies cover the application layer, as explained, the proposed adaptations are quite limited, showing that composition is still challenging, particularly when considering two specific quality attributes: resilience and security.

Resilience is related to the fact that the application is able to recover from a failure and continues its execution. Those failures can be caused by different factors, some due to the very distributed nature of microservices, such as unavailability and overload, and others related to the composition itself, such as human errors during the microservice implementation and the creation of the composition flow. Therefore, the recovery of failures can be different according to the criticality level that the activity implemented by the microservice has on the composition process. In this sense, 
although resilience is the third most addressed quality attribute, most of the studies focus on the infrastructure layer, which indicates a lack of work that propose self-adaptive approaches for resilient microservice composition in the application layer.

Security and reliability are another critical factors, since an application based on microservices can modify its structure continuously during execution due to the dynamic aspect of the microservices. In addition, as the microservice architecture provides flexibility to develop microservices using diverse technologies, an application can consist of heterogeneous microservices that use different operating systems, programming languages, data stores and other auxiliary tools. As a result, this technological diversity increases the possibility of attacks that can exploit specific security vulnerabilities. Therefore, during composition, a particular microservice that is needed to perform a  distributed business process can be compromised by an internal (for example, an unauthorized person who interferes with the production environment) or external (hackers who deliberately try to make the application's service unavailable through attacks such as Denial of Service (DoS). As a result, the entire business process is also compromised. 

Table~\ref{tab:table_concerns_addressed} shows that those two quality attributes are the focus of only two studies each, which demonstrates that there is a gap on self-adaptive approaches for dealing with security and reliability. Furthermore, Table~\ref{tab:table_self_properties} confirms that self-protecting is currently not addressed by any study, thus corroborating with the importance of such research direction.

\subsection{Languages and Models for Adaptation Planning}

An important approach to self-adaptation depends mainly on architectural models that provide a high-level view of a system and allow to detect when a system fails to meet its objectives \cite{cheng2012}. There are several works in the literature that address self-adaptation techniques through architectural models, associated with domain-specific languages (DSL), such as the Stitch language in \cite{cheng2012}. Despite showing great potential and being used in other contexts of self-adaptation, mainly in the creation of adaptation plans, this approach is almost unexplored in the context of microservices. According to Figure~\ref{fig:pesquisa_vs_contribuicao}, only one study proposed languages to deal with self-adaption on microservices. This gap is also reflected by Table~\ref{tab:table_overview_techonologies}, which indicates that modeling tools and specification languages are addressed by only one study each. 

In this sense, a promising approach consists on investigating the use of DSLs that can assist the task of specifying the possible adaptation actions using a higher abstraction level in the context of self-adaptive microservices. For instance, after the need to adapt the microservice-based application is detected, it is necessary to specify how the adaptation will occur. The adaptation may be related either to infrastructure elements, such as creating new containers for an overloaded microservice, or to the execution flow of a microservice composition, such as retrying to perform the service or canceling the whole composition process. 

\subsection{DevOps and Self-adaptive deployment}

With the increasing adoption of DevOps tools and practices by companies, such as continuous integration and continuous delivery, aiming at developing agile cloud-native systems, the deployment of new application updates or microservices in production becomes a challenge. Nonetheless, according to Table~\ref{tab:table_overview_techonologies}, only 3 studies addressed DevOps technologies for providing self-adaptation capabilities for microservice-based systems. In [S1] the authors explored the deployment of microservices autonomously in order to improve their allocation based on affinity and in [S17] the authors proposed a new approach to manage different versions of microservices without human intervention. However, no study took into account the deployment of microservice updates in a progressive manner and with the least possible impact. Thus, an interesting possibility would be to explore the automation of deployments, for example, canary deployment\footnote{https://martinfowler.com/bliki/CanaryRelease.html}, which requires a high level of monitoring of what should be implemented, as an activity that involves the use of self-adaptation techniques through feedback control loops. This could open up a range of possibilities for new self-adaptive systems for microservices, from systems that provide basic self-scaling to systems that provide self-adaptive deployments.

\subsection{Microservices and machine learning}

Although machine learning (ML) techniques are being used for a variety of  application domains, its use in self-adaptive microservice-based systems is still in early stages. In fact, as depicted by Table~\ref{tab:table_overview_techonologies}, only 3 studies used ML-based approaches. In this regard, there is a great opportunity to add intelligence to different levels of microservice applications. For instance, ML can be used to implement more proactive adaptation strategies, which would anticipate possible problems in the system, such as battery discharging, as addressed by [S19]. According to Table~\ref{tab:table_adaptation_strategy}, such kind of adaptation strategy has been proposed by only one study. In addition, ML techniques could pave the way to develop more decentralized adaptation solutions that intelligently monitor the microservices communication and adjust the exchanging message flow to avoid bottlenecks. At last, but not least, ML algorithms can be very useful to implement self-adaptation approaches to deal with uncertainties in microservice applications. For instance, learning unexpected events that may happen in the environment at runtime can turn the application more robust and resilient.    

\section{Threats to Validity}
\label{sec:threats-to-validaty}

Although our study was carried out using a strict protocol in accordance with the guidelines well established in the literature \cite{kitchenham2007} \cite{kitchenhambook2015} \cite{zhanghe2011}, it has potential threats to validity that will be discussed in the following paragraphs along with the strategies we have adopted to mitigate them.

\textbf{\textit{Construct Validity}}. In our study, we used five most important electronic databases in the field of computer science in order to select a large number of relevant papers. Despite this, it is still possible that some studies have not been returned by the selected databases. One way found to mitigate this threat was to carry out a snowball procedure (backward and forward), as presented in Section \ref{sec:planning-the-mapping}. In relation to threats belonging to a malformation of the search string, we compose two groups of terms related to microservices and self-adaptation, formed by keywords based on the main publications for each context. In addition, we refined our search string by conducting pilot surveys.

\textbf{\textit{Internal Validity}}. In the data extraction process, the same data item can be interpreted differently by each researcher involved in the process, leading to divergent data. We have mitigated this threat by establishing base references for each data item to be extracted. Additionally, all the data items to be extracted were shared in a spreadsheet that made the researchers' understanding clearer.

\textbf{\textit{External Validity}}. Regarding the ability to generalize the conclusions, usually related to the representativeness of the sample, approaches based on techniques and mechanisms of self-adaptation in the architectural style of microservices were considered. Therefore, the classification presented and the conclusion drawn are valid only in this context.

\textbf{\textit{Conclusion Validity}}. To obtain a systematic quality study, it is essential to have a reliable and well-defined study classification scheme. Still, this is another point where subjectivity can bring threats. To mitigate this problem, the entire process of classifying the studies was carried out in pairs, as described in Sections \ref{sec:planning-the-mapping} and \ref{sec:conducting-the-mapping}, in addition to discussions and meetings held in cases of disagreement, aiming to reach a sense.

\section{Final Remarks}
\label{sec:conclusion-and-future-works}

In this paper, we have conducted a systematic mapping that, through 10 primary research questions, identified how self-adaptation approaches are being applied to microservice-based architectures. The paper selection occurred between February 2020 and May 2020, resulting in 21  studies that were analyzed. Our selection process happened using an automatic search in five scientific databases and the snowballing technique.

The main findings of this work are: (i) most of the studies aimed at developing solutions that deal with self-healing and addressed mainly the performance and scalability quality attributes; (ii) reactive adaptation strategies, through a centralized adaptation control logic, and in the system infrastructure level, were the focus of the majority of the studies; and (iii) most of the analyzed studies proposed an architectural model that addresses self-adaptive capabilities in microservice-based applications. By analyzing the systematic mapping results, it was possible to propose four promising research directions that have not been properly addressed yet. 

In future work, we intend to explore other results of the systematic mapping that could not be covered in this paper. We also plan to research the topics proposed in our roadmap. Finally, we plan to investigate the use of self-adaptive approaches in more recent service-based architectures, such as serverless and function as a service.

\section{Acknowledgments}

This work is partially supported by CNPq/Brazil under grant Universal 438783/2018-2.


\bibliographystyle{IEEEtran}
\bibliography{main}

\vspace{12pt}

\end{document}